\input ./style/arxiv-general.cfg
\documentclass[MSNbibl,number,dvips]{arxstspdf}
\makeatletter
   \@ifpackageloaded{graphicx}{}{\usepackage{graphicx}}
\makeatother
\usepackage{stfloats}
\usepackage{flushend}
\usepackage{stfloats}

\volume{30}
\issue{2}
\pubyear{2015}
\firstpage{167}
\lastpage{169}
\doi{10.1214/15-STS516}
\referstodoi{10.1214/00-STSXXXX}
\docsubty{FLA}

\makeatletter
  \fnbelowfloat
\newcommand{\rrvert}{\vert}
\newcommand{\rrVert}{\Vert}
\newcommand{\llvert}{\vert}
\newcommand{\llVert}{\Vert}

\def\bC{\mathbf{C}}
\def\btheta{\bolds{\theta}}
\def\R{\mathbb{R}}
\def\bh{\mathbf{h}}
\makeatother

\begin{document}
\begin{frontmatter}
\vspace*{12pt}
\title{On the Flexibility of Multivariate Covariance Models:
Comment on the~Paper by Genton and Kleiber}
\referstodoi{10.1214/14-STS487}
\runtitle{Comment}

\begin{aug}
\author[A]{\fnms{Moreno}~\snm{Bevilacqua}\ead[label=e1]{moreno.bevilacqua@uv.cl}},
\author[B]{\fnms{Amanda S.}~\snm{Hering}\ead[label=e2]{ahering@mines.edu}}
\and
\author[C]{\fnms{Emilio}~\snm{Porcu}\corref{}\ead[label=e3]{emilio.porcu@usm.cl}}
\runauthor{M. Bevilacqua, A.~S. Hering and E. Porcu}

\affiliation{University of Valparaiso, Colorado School of Mines and Technical University Federico Santa Maria}

\address[A]{Moreno Bevilacqua is Associate Professor, Department of Statistics, University of Valparaiso, Chile \printead{e1}.}
\address[B]{Amanda S. Hering is Assistant Professor,
Department of Applied Mathematics and Statistics, Colorado School of Mines,
Golden, Colorado 80401, USA \printead{e2}.}
\address[C]{Emilio~Porcu is Professor, Department of Mathematics, Technical University Federico Santa Maria, Valparaiso, Chile \printead{e3}.}
\end{aug}


\end{frontmatter}

\section*{Introduction}

We congratulate the authors for their considerable effort to collect
and synthesize all of the information contained in this review paper.
Given the breadth of models, we were particularly inspired by the idea
of how a practitioner would choose among them.
We define some general criteria of flexibility that should be
considered when choosing between different multivariate covariance
models, and we apply these criteria
in the comparison between the bivariate linear model of
coregionalization (LMC) and the bivariate multivariate Mat\'{e}rn.

\subsection*{Which Model Is the Most Flexible?}

Since most of the contributions listed by the authors refer to
parametric models of multivariate covariances, we seek to answer the
question, ``which parametric model is more flexible?'' We propose to
define flexibility with respect to the following:
\begin{longlist}[(A)]
\item[(A)] the colocated correlation coefficient, and
\item[(B)] the strength of spatial dependence. For instance, how
different can the scales of the cross-covariances and the marginal
covariances between the two models be.
\end{longlist}
As far as (A) is concerned, ideally the colocated correlation
coefficient should be defined over the interval $[-1,1]$. Let us
consider models of the type
%
\begin{equation}
\label{eqma} \quad\bC(\bh)= \bigl[ \sigma_i \sigma_j
\rho_{ij} R(\bh;\btheta_{ij} ) \bigr]_{i,j=1}^2,
\qquad\bh\in\R^d,
\end{equation}
with $R(\cdot)$ being a parametric univariate correlation model in $\R
^d$ and $\btheta_{ij} \in A \subset \R^q$ being parameter vectors.
Here $\sigma_i^2>0$, $i=1,2$ are the marginal variances, and $\rho
_{12}$ is the colocated parameter describing the correlation between
the components of the bivariate random field at $\bh=0$.
The bivariate Mat{\'e}rn \cite{GKS} and Wendland \cite{PDB} models are
special cases of equation~(\ref{eqma}).

\begin{figure}

\includegraphics{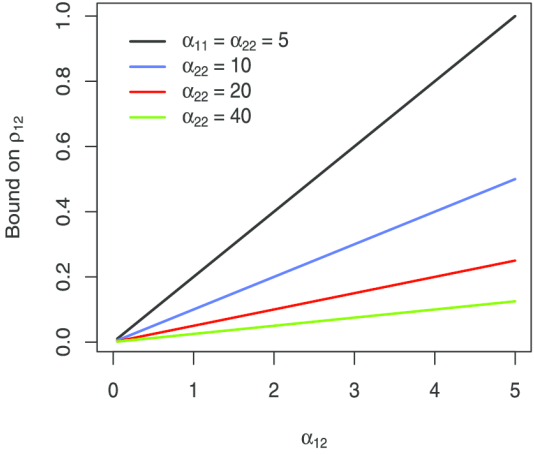}

\caption{The upper bound on $\rho_{12}$ as a function of $\alpha_{12}$
for various values of $\alpha_{11}$ and $\alpha_{22}$. Note that for
the colored lines, $\alpha_{11}=5$.}\label{figboli}
\end{figure}

For the bivariate Mat{\'e}rn case, the validity bound for $\rho_{12}$
is given in Theorem 3 of \cite{GKS},
and in general it depends on the smoothness parameters, $
\bolds{\nu}=(\nu_{11},\nu_{22}, \nu_{12})'$, and the scale
parameters, $ \bolds{\alpha}=(\alpha_{11},\alpha
_{22},\alpha_{12})'$. For instance, assuming a constant smoothness
parameter, and $ \alpha_{12}<\min( \alpha_{11}, \alpha_{22})$, a
necessary and sufficient condition for the validity of the bivariate
Mat{\'e}rn becomes $|\rho_{12}| \leq\frac{\alpha^2_{12}}{ \alpha_{22}
\alpha_{11}} \leq1$.
In\vspace*{1pt} the case where the scale and smoothness parameters are pairwise
equal (i.e., the separable case), then $| \rho_{12}| \leq1$, and there
are no restrictions on the colocated parameter. These features are also
present in the bivariate Wendland construction in \cite{PDB}, where the
elements of the matrix-valued covariance are parameterized in the same
way as the bivariate Mat{\'e}rn. As the difference between the
parameters $\alpha_{11}$ and $\alpha_{22}$ increases, the bound on $\rho
_{12}$ becomes tighter, as shown in Figure~\ref{figboli}.

The linear model of coregionalization (LMC) does not necessarily share
this limitation on the colocated correlation coefficient. In order to
illustrate this, we start with a simple example: 
%
%
for the following, we write $R(\cdot):= C(\cdot)/C(0)$, for $C$ some
univariate covariance function in $\R^d$. Then, the bivariate LMC
correlation model $\mathbf{R}(\bh)= [ R_{ij}(\bh) ]_{i,j=1}^2$,
$\bh\in\R^d$, can be written as
\begin{eqnarray*}
R_{11}(\bh) &=& a_{11}^2 R_1(\bh)+
a_{12}^2 R_2(\bh),
\\
R_{22}(\bh) &=& a_{21}^2 R_1(\bh)+
a_{22}^2 R_2(\bh),\quad\mbox{and}
\\
R_{12}(\bh) &=& a_{11}a_{21} R_1(
\bh)+ a_{12}a_{22} R_2(\bh).
\end{eqnarray*}
The $2\times2$ matrix $\mathbf{A}=\{a_{ij}\}$ has rank 2.
Here we focus, without loss of generality, only on positive\vadjust{\goodbreak}
correlations between the components, and, in order to do that, we
consider $a_{ij}>0$ for $i,j=1,2$.
Let us now consider the special case $a_{12}=a_{21}$. Since
$R_{ii}(\mathbf{0})$ must be\vspace*{1pt} identically equal to one, we get, as
necessary conditions, that $a_{11}^2+a_{12}^2=1=a_{22}^2+a_{12}^2$,
which in turn implies that necessarily $a_{11}=a_{22}=:a$. Then the
previous system of equations can be rewritten as
\begin{eqnarray*}
R_{11}(\bh)&=& a^2 R_1(\bh)+
\bigl(1-a^2\bigr) R_2(\bh),
\\
R_{22}(\bh)&=& \bigl(1-a^2\bigr) R_1(\bh)+
a^2 R_2(\bh),\quad\mbox{and}
\\
R_{12}(\bh) &=& a\sqrt{1-a^2} \bigl( R_1(
\bh)+ R_2(\bh) \bigr).
\end{eqnarray*}
In this case the colocated correlation coefficient is identically equal
to $\rho_{12}:= R_{12}(\mathbf{0})=2 a \sqrt{1-a^2} \in[0,1]$
since $a\in[0,1]$. Thus, the LMC seems to be more flexible than the
bivariate Mat\'{e}rn
with respect to issue (A) since its colocated correlation coefficient
is free to vary through the maximum extent of its possible range,
regardless of the form of the marginal covariances.

%

Issue (B) is clearly more critical to address, since it is not directly
interpretable from the parameterization. In particular, it would be
nice to have models that allow for different levels of strength of
spatial correlation, which is directly related to the scales. Clearly,
the conditions on the bivariate Mat{\'e}rn as well as those on the
bivariate Wendland indicate that we have an ill-posed problem because
the colocated correlation coefficient's upper bound is related to the
scales. Thus, we analyze issue (B) for a fixed value of $\rho_{12}$. In
particular, we try to address the question: ``which model allows for
bigger differences with respect to the strength of spatial\vadjust{\goodbreak} correlation
for a given colocated correlation coefficient?''
We introduce here two multivariate measures that once again we
illustrate for the bivariate case for ease of exposition.

[(B.1)] Since we are dealing with isotropic models, we shall write
$R(t)$ instead of $R(\|\bh\|)$ for $\bh\in\R^d$. For a given
multivariate covariance model $x$, we define
%
\begin{equation}
\label{bidoug} {\mathcal D}^{x}_{i,j,k}:= \max
_{t} \bigl\llvert R_{ii}(t)-R_{kj}(t)
\bigr\rrvert,\quad t:= \llVert\bh\rrVert>0,\hspace*{-20pt}
\end{equation}
where\vspace*{1pt} $k \in\{i,j\}$.
When\vspace*{1pt} $k=j\neq i$,
${\mathcal D}^{x}_{i,j,j}={\mathcal D}^{x}_{j,i,i}$ is a measure of the
maximum difference between the correlation of the $i$th and $j$th components;
while for $k=i$, ${\mathcal D}^{x}_{i,j,i}={\mathcal D}^{x}_{i,i,j}$
reflects\vspace*{1pt} the maximum difference between the cross correlation $R_{ij}$
and the correlation of the $i$th component.
According to this criterion, for two given models $x$ and $y$, with a
common (fixed) colocated correlation coefficient $\rho_{12}^x=\rho
_{12}^y$, if ${\mathcal D}^{x}_{i,k,j}<{\mathcal D}^{y}_{i,k,j}$, then
model $y$ is preferable.\vspace*{1pt}
%

The computation of the indicator above can be tedious, depending on the
functional forms of the involved marginal and cross correlations.
For instance, for the bivariate Mat{\'e}rn model, obtained when fixing
the parameters $\nu_{11}$, $\nu_{22}$ and $\nu_{12}$ to be identically
equal to $0.5$, the general form of the bivariate correlation function
can be written as $0 \le t \mapsto R_{ij}(t)=\rho_{ij} e^{-t\alpha
_{ij}}$, $\alpha_{ij}>0$, $\rho_{ii}=1$.
In this case, inspection of equation~(\ref{bidoug}) directly relates to
finding the stationary solution of the problem
\[
G(a,b,\rho):=\max_{t \ge0} \bigl\llvert f(t,a,b,\rho) \bigr
\rrvert, %
\]
where $f(t,a,b,\rho)= {\rm e}^{-at}- \rho{\rm e}^{-bt}$, with $a,b>0$,
and $\rho$ could be a function of $a$ and $b$ but is fixed here and
belongs to the interval $[-1,1]$. The case $\rho<0$ is trivial since
the maximum is attained at $t=0$, so we focus on the case $\rho>0$.
The problem has the following solutions:
%
\begin{equation}
\label{eqG} G(a,b,\rho)= \cases{ \displaystyle \max \bigl(-f \bigl(t^*,a,b,
\rho \bigr),1- \rho \bigr),
\vspace*{3pt}\cr
\quad \mbox{if $b<a$},
\vspace*{3pt}\cr
f \bigl(t^*,a,b,\rho
\bigr),
\vspace*{3pt}\cr
\quad \mbox{if $b>a$, $ \log\rho+ \log(b/a)>0$,}
\vspace*{3pt}\cr
1-\rho,
\vspace*{3pt}\cr
\quad
\mbox{if $b>a$, $\log\rho+\log(b/a)<0$,}}\hspace*{-23pt}
\end{equation}
where $t^*=\frac{\log\rho+\log(b/a)}{b-a}$. For $b=a$, $ G(a,b,\rho
)= 1-\rho$.
For the bivariate exponential model, we can\vadjust{\goodbreak} compute
\begin{eqnarray*}
{\mathcal D}_{1,2,2}^{\mathrm{Exp}} &=& G(\alpha_{11},
\alpha_{22},1)
\\
&=& \cases{ 0,\quad \mbox{$\alpha_{11}=\alpha_{22}$},
\vspace*{3pt}\cr
\displaystyle -f \biggl(\frac{\log(\alpha_{22}/\alpha_{11})}{\alpha_{22}-\alpha_{11}},\alpha _{11},
\alpha_{22},1 \biggr),
\vspace*{3pt}\cr
\hspace*{22pt} \mbox{if $\alpha_{22} <\alpha_{11}$},
\vspace*{3pt}\cr
\displaystyle f \biggl(\frac{\log(\alpha_{22}/\alpha_{11})}{\alpha_{22}-\alpha_{11}},\alpha
_{11},\alpha_{22},1 \biggr),
\vspace*{3pt}\cr
\hspace*{22pt}\mbox{if $\alpha_{22}>\alpha_{11}$}.}
\end{eqnarray*}
Similarly, ${\mathcal D}_{1,2,1}^{\mathrm{Exp}}$ and ${\mathcal
D}_{2,2,1}^{\mathrm{Exp}}$ can be computed using equation (\ref{eqG}) as
$G(\alpha_{11},\alpha_{12},\rho_{12})$ and
$G(\alpha_{22},\alpha_{12},\rho_{12})$.
Note\vspace*{1pt} that ${\mathcal D}^{\mathrm{Exp}}_{1,2,2}$ does\vspace*{1pt} not depend on the
colocated correlation coefficient and that, for this example, $0\leq
{\mathcal D}^{\mathrm{Exp}}_{1,2,2} \leq1$, and, similarly, we have $0\leq
{\mathcal D}^{\mathrm{Exp}}_{1,2,1} \leq1$.\vspace*{1pt}

Let us now analyze a special case of the LMC model as illustrated
through issue (A), supposing that $R_{1}(t)= \exp(-\alpha t)$ and
$R_{2}(t)= \exp(-\beta t)$, for\vspace*{1pt} $\alpha$ and $\beta$ positive.
Direct inspection shows that in this case ${\mathcal D}^{\mathrm{LMC}}_{1,2,2}=
k | 2a^2 -1 | $,
where $k= G(\alpha,\beta,1)$.
Now, we note that $\rho_{12}=2 a \sqrt{1-a^2}$ for $0<a<1$ and, as
shown before, $\rho_{12}$ can belong to any value inside the interval
$[0,1]$. Since
$a= [\frac{1 \pm\sqrt{1-\rho_{12}^2}}{2} ]^{0.5}$, then ${\mathcal
D}^{\mathrm{LMC}}_{1,2,2}=k\sqrt{1-\rho_{12}^2}\le\sqrt{1-\rho_{12}^2}$, with
equality if and only if $\rho_{12}=0$. Similarly, it can be shown that
${\mathcal D}^{\mathrm{LMC}}_{1,2,1} \le1-\rho_{12}$ and ${\mathcal
D}^{\mathrm{LMC}}_{2,2,1} \le1-\rho_{12}$.\vspace*{1pt}

Comparing the index between the bivariate LMC and exponential model,
when $\rho_{12} \to1$, then ${\mathcal D}^{\mathrm{LMC}}_{1,2,2}< {\mathcal
D}^{\mathrm{Exp}}_{1,2,2}$. Thus, it seems that more flexibility is offered by
the bivariate exponential at least for the marginal
correlations $R_{11}$ and $R_{22}$.

[(B.2)] As a second indicator for a given multivariate covariance model
$x$, we propose
%
\begin{equation}
\label{bidoug2} \widetilde{{\mathcal D}}^x_{i,k,j}:= \biggl
\llvert \int_{0}^{\infty} \bigl( R_{ii}(t) -
R_{kj}(t) \bigr)\, {\rm d} t \biggr\rrvert .
\end{equation}
For instance, let us consider the case of the radial part of a
bivariate Mat{\'e}rn model, so that
\[
R_{ij}(t)= \rho_{ij} (\alpha_{ij}
t)^{\nu_{ij}} {\mathcal K}_{\nu
_{ij}}(\alpha_{ij}t), \qquad t
\ge0, %
\]
for which direct inspection shows that
\begin{eqnarray*}
\widetilde{{\mathcal D}}^{\mathrm{Mat}}_{1,1,2} &=& \biggl\llvert \sqrt{
\pi} \biggl(\frac
{\Gamma (\nu_{11}+(1/2) )}{\alpha_{11} \Gamma(\nu_{11})}
\\
&&\hspace*{2pt}{} -\frac
{\rho_{12} \Gamma (\nu_{12}+(1/2) )}{\alpha_{12} \Gamma(\nu
_{12})} \biggr) \biggr\rrvert .
\end{eqnarray*}
Observe that in a bivariate exponential model, we easily get $\widetilde
{{\mathcal D}}^{\mathrm{Exp}}_{1,1,2} = \llvert  \alpha_{11}^{-1} - \rho_{12}
\alpha_{12}^{-1} \rrvert  $,
and $\widetilde{{\mathcal D}}^{\mathrm{Exp}}_{1,2,2}=\llvert  \alpha
_{11}^{-1} - \alpha_{22}^{-1} \rrvert  $ does\vspace*{1pt} not depend on the
colocated correlation coefficient (as in the previous index).
For the LMC model we easily get that $\widetilde{{\mathcal
D}}^{\mathrm{LMC}}_{1,2,2} = \sqrt{1-\rho_{12}^2} \llvert  \alpha^{-1} - \beta
^{-1} \rrvert $ for the exponentials as assumed earlier.
As with the first index when $\rho_{12} \to1$, $\widetilde{\mathcal
D}^{\mathrm{LMC}}_{1,2,2}<\widetilde{\mathcal D}^{\mathrm{Exp}}_{1,2,2}$. Again\vspace*{1pt} the
bivariate exponential model seems to be more flexible with respect to
LMC, at least when comparing the marginal correlations $R_{11}$ and $R_{22}$.\vspace*{10pt}

\section*{Conclusion}
Comparing multivariate covariance models from a flexibility point of
view is an important issue, since it can be helpful when choosing
between the set of parametric models
described in this review paper.
In this discussion we have proposed two possible criteria in order to
define the flexibility of a multivariate covariance model. Using the
criteria proposed, we compare the flexibility of the LMC with the
multivariate Mat{\'e}rn in the bivariate case. There is a clear
trade-off between the two models. In particular, the LMC is more
flexible in terms of the allowable interval for the colocated
correlation coefficient, which varies freely over the interval
$[-1,1]$. Instead, the bivariate Mat{\'e}rn has restrictions on the
permissible range of the colocated correlation for given spatial scales
and smoothness parameters.
On the other hand, the two indices highlight that if the colocated
correlation is the same for the two models, the bivariate Mat{\'e}rn
has more flexibility as the colocated correlation tends to 1.

In conclusion, we believe that more work should be done in order to
compare multivariate models from a flexibility viewpoint. In this
discussion, we have offered some instruments that might be useful in
this direction.




%

\end{document}